\documentclass[aps,prl,twocolumn,amsmath,amssymbs]{revtex4-2}

\usepackage{graphicx}
\usepackage{amsmath,amsfonts,amssymb}
\usepackage{epsfig}
\usepackage{wrapfig}
\usepackage{bbm}
\usepackage[usenames]{color}
\usepackage{array}
\usepackage{times}

\newcommand{\be}{\begin{equation}}
\newcommand{\ee}{\end{equation}}

\def\W{\omega}
\def\um{\mu\text{m}}

\def\dW{\delta\omega}

\begin{document}


\title{Tunable quantum interference using a topological source of indistinguishable photon pairs}

\author{Sunil Mittal$^{1,\,2,\,*,\,\dag}$}
\author{Venkata Vikram Orre$^{1,\,2,\,*}$}
\author{Elizabeth A. Goldschmidt$^{3}$}
\author{Mohammad Hafezi$^{1,\,2,\,4}$}

\affiliation{$^1$Joint Quantum Institute, NIST/University of Maryland, College Park, Maryland 20742, USA}
\affiliation{$^2$Department of Electrical and Computer Engineering and IREAP, University of Maryland, College Park, Maryland 20742, USA}
\affiliation{$^3$Department of Physics, University of Illinois at Urbana-Champaign, Urbana, Illinois 61801, USA}
\affiliation{$^4$Department of Physics, University of Maryland, College Park, MD 20742, USA}
\affiliation{$^*$S.M. and V.V.O. contributed equally}
\affiliation{$^\dag$Email: mittals@umd.edu}

\begin{abstract}
Sources of quantum light, in particular correlated photon pairs that are indistinguishable in all degrees of freedom, are the fundamental resource that enables continuous-variable quantum computation and paradigms such as Gaussian boson sampling. Nanophotonic systems offer a scalable platform for implementing sources of indistinguishable correlated photon pairs. However, such sources have so far relied on the use of a single component, such as a single waveguide or a ring resonator, which offers limited ability to tune the spectral and temporal correlations between photons. Here, we demonstrate the use of a topological photonic system comprising a two-dimensional array of ring resonators to generate indistinguishable photon pairs with dynamically tunable spectral and temporal correlations. Specifically, we realize dual-pump spontaneous four-wave mixing in this array of silicon ring resonators that exhibits topological edge states. We show that the linear dispersion of the edge states over a broad bandwidth allows us to tune the correlations, and therefore, quantum interference between photons by simply tuning the two pump frequencies in the edge band. Furthermore, we demonstrate energy-time entanglement between generated photons. We also show that our topological source is inherently protected against fabrication disorders. Our results pave the way for scalable and tunable sources of squeezed light that are indispensable for quantum information processing using continuous variables.
\end{abstract}

\maketitle

Spurred by the possibility of realizing continuous-variable quantum computation, and gaussian boson sampling that holds near-term promise for quantum simulations and various graph-theory problems, nanophotonics systems have emerged as a natural platform to generate indistinguishable correlated photon pairs and, in the strong nonlinearity regime, single-mode squeezed light \cite{Braunstein2005, Pfister2019, Hamilton2017, Vernon2019, Silverstone2014, He2015, Zhao2020, Zhang2020}. Most on-chip sources of indistinguishable photon pairs have relied on dual-pump spontaneous four-wave mixing (DP-SFWM), a third-order nonlinear process, in silicon or silicon-nitride waveguides and ring resonators \cite{Silverstone2014, He2015, Zhao2020, Zhang2020, Vernon2019}. In this process, two pump photons at different frequencies annihilate and create two frequency-degenerate photons, called signal and idler. The tight mode confinement in nanophotonic waveguides and the use of a ring resonator remarkably enhance the strength of SFWM interactions, leading to an enhancement in pair generation rates and eventually on-chip sources of squeezed light \cite{He2015, Dutt2015, Zhao2020, Zhang2020}. Extensions of these simple single-element systems to multi-mode, multi-resonator systems can allow tunability and multiplexing of various spectral or temporal modes, and thereby, enable a significant reduction in physical resources \cite{Pfister2019, Takeda2019, Asavanant2019, Larsen2019}. However, such extensions have so far remained elusive.

At the same time, the influx of ideas derived from the physics of topological insulators has led to a new paradigm of complex photonic devices that use arrays of coupled waveguides or resonators to achieve unprecedented control over the flow of photons \cite{Lu2014, Khanikaev2017, Ozawa2019, Haldane2008, Wang2009, Hafezi2011, Kraus2012, Hafezi2013, Rechtsman2013}. More specifically, edge states, the hallmark of topological systems, exhibit unique features such as unidirectional (or helical) flow of photons confined to the boundaries of a system, linear dispersion, and an inherent robustness against disorders. Photonic edge states have now been used to realize robust optical delay lines \cite{Hafezi2011, Hafezi2013, Mittal2014}, lasers \cite{St-Jean2017, Bahari2017, Bandres2018}, optical fibers \cite{Lu2018}, and reconfigurable pathways on chips \cite{Cheng2016, Zhao2019}. More recently, topological edge states have been exploited in quantum photonic devices, to realize chiral quantum-optic interfaces between quantum dots and photonic crystals \cite{Barik2018}, topological beamsplitter for quantum interference of photons \cite{Tambasco2018}, and also topological sources of quantum light \cite{Mittal2018, Blanco-Redondo2018}. In particular, in ref.\cite{Mittal2018} we implemented a topological source of distinguishable photon pairs using single-pump SFWM in a two-dimensional (2D) lattice of coupled ring resonators. Therefore, it is intriguing to investigate photonic designs inspired by topological physics for the development of tunable and robust sources of indistinguishable photon pairs.

\begin{figure*}[!ht]
\centering
\includegraphics[width=0.9\textwidth]{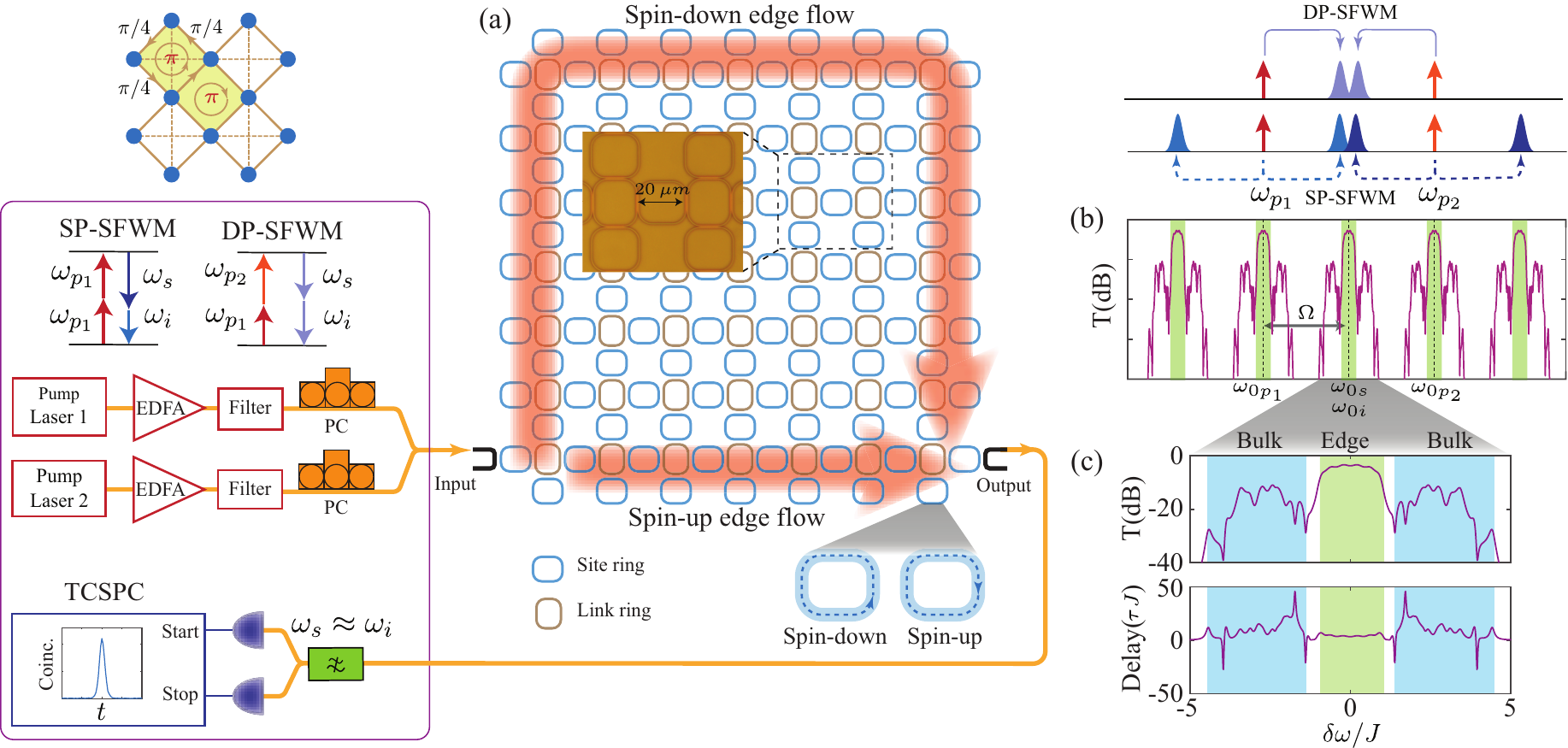}
\caption{
(a) Schematic of the 2D array of silicon ring resonators that simulates the anomalous quantum-hall effect for photons. The link rings (shaded brown) couple nearest and next-nearest neighbor site rings (shaded blue), with hopping phases as shown in the inset. The lattice supports two pseudospins (up and down), with corresponding edge states travelling in opposite directions. (b,c) Simulated transmission and delay spectra of the device showing edge and bulk bands. The spectra repeats after every FSR. To generate indistinguishable photon pairs via dual-pump SFWM, the lattice is pumped using two continuous-wave lasers, in different FSRs centered at $\W_{0p_{1}}$, and $\W_{0p_{1}}$ (top-right inset). At the output, we use superconducting nanowire detectors (SNSPDs), tunable filters centered around $\W_{s} \approx \W_{i}$, and time-resolved coincidence detection. This measurement setup setup allows us to exclude frequency non-degenerate photons created by single pump SFWM (top-right inset).  EDFA: erbium-doped fiber amplifier; PC: polarization controller; TCSPC:time-correlated single photon counter.
}
\label{fig:1}
\end{figure*}

In this article, we report the generation of indistinguishable photon pairs via dual-pump SFWM in a 2D lattice of coupled ring resonators. This lattice realizes the anomalous quantum Hall model for photons and exhibits topological edge states \cite{Haldane1988, Leykam2018, Mittal2019, Hafezi2013}. We show that the linear dispersion of the edge states results in phase-matched generation of photon pairs throughout the edge band, and therefore, allow us to tune the spectral-temporal bandwidth of photon pairs by tuning the input pump frequencies in the edge band. To show that the generated photon pairs are indeed indistinguishable, we use the fact that our systems is time-reversal symmetric and therefore, supports two pseudo-spins that circulate around the lattice in opposite directions (Fig.\ref{fig:1}). We use these counter-propagating edge states to realize a Sagnac interferometer and deterministically split the indistinguishable photon pairs \cite{Chen2007, Silverstone2014, He2015}. We then demonstrate Hong-Ou-Mandel (HOM) interference between split photons which unequivocally establishes their indistinguishability. Furthermore, we show that the tunability of the spectral bandwidth of our source manifests in the temporal width of the HOM interference dip. Finally, we demonstrate that the generated photon pairs are energy-time entangled \cite{Franson1989} as expected for a SFWM process with a continuous-wave pump, and our source is robust against fabrication disorders. Our results could lead to the realization of on-chip, topologically robust, and spectrally engineered sources of squeezed light for applications in continuous-variable quantum computation and gaussian boson sampling \cite{Pfister2019, Braunstein2005, Hamilton2017}.

\begin{figure*}[!ht]
\centering
\includegraphics[width=0.9\textwidth]{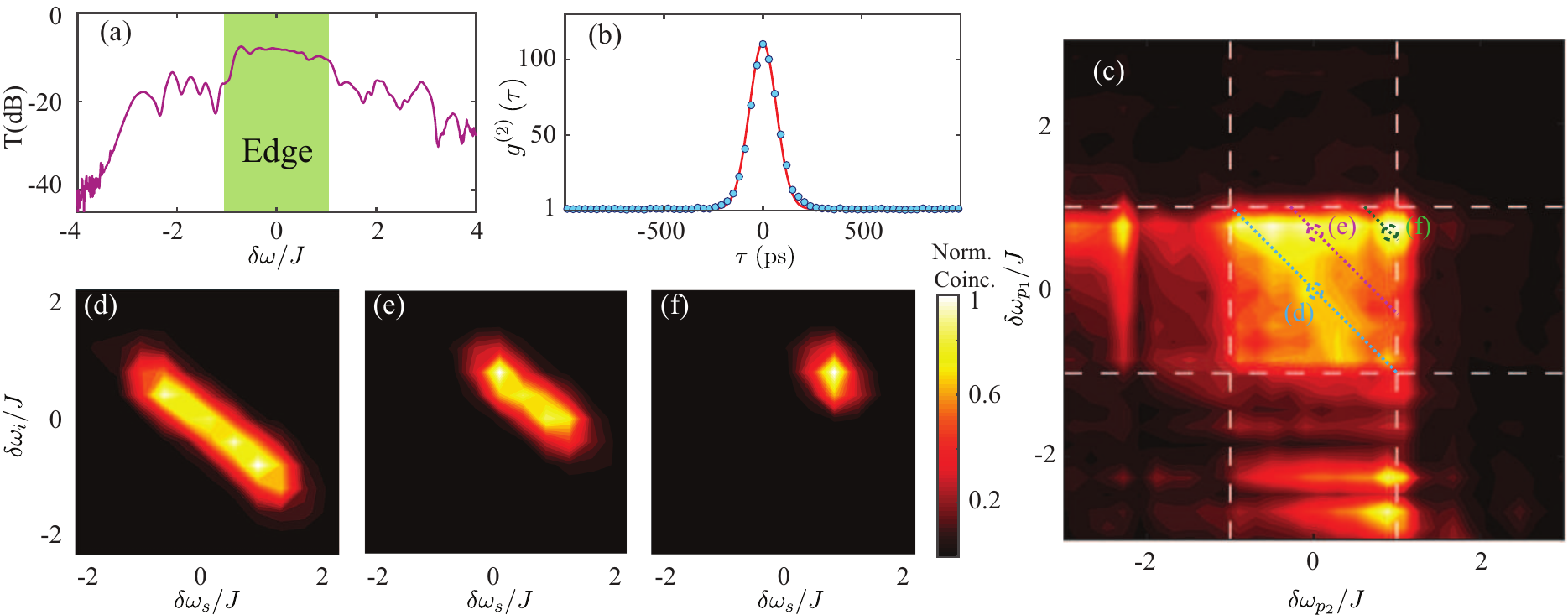}
\caption{
(a) Measured transmission spectrum of an anomalous Hall device, with the edge band highlighted in green. (b) Measured second-order temporal correlation function $g_{2}\left(\tau \right)$ showing that the generated signal and idler photons are strongly correlated. (c) Measured coincidence counts as function of the input pump frequency detunings $\dW_{p_{1}},\dW_{p_{2}}$ relative to the respective longitudinal mode resonances. (d-f) Measured joint spectral intensity of the signal and idler photons, for input pump frequencies $\left(\dW_{p_{1}}, \dW_{p_{2}}\right)$ = $\left(0, 0 \right)$, $\left(0.8, 0 \right) J$, and $\left( 0.8, 0.8 \right) J$. The pump frequencies are also indicated in (c). The dashed lines show the spectra of generated photons in the edge band allowed by energy conservation.
}
\label{fig:2}
\end{figure*}

Our system consists of 2D checkerboard lattice of ring resonators (Fig.\ref{fig:1}(a)) \cite{Leykam2018, Mittal2019}. The rings (shaded blue) at the lattice sites are coupled to their nearest and next-nearest neighbors using another set of rings, which we call the link rings \cite{Hafezi2011, Hafezi2013}. The gap between the link and the site rings sets the strength $\left(J \right)$ of the evanescent field coupling between the site rings and is the same for both the nearest and next-nearest neighbor site rings. The resonance frequencies of the link rings are detuned from those of the site rings such that the link rings act as waveguides connecting site rings. More importantly, depending on their position, the link rings introduce a direction dependent hopping phase between the site rings. In our system, the link rings are positioned such that the hopping phases between next-nearest neighbor site rings is always zero, and that between nearest neighbor site rings is $\pm \pi/4$. This configuration effectively leads to the realization of a staggered synthetic magnetic field for photons such that the average magnetic flux through a unit cell of two plaquettes of the lattice is zero (shaded yellow in Fig.\ref{fig:1}a), but the flux through a single plaquette is non-zero. This coupled ring resonator configuration simulates the anomalous quantum Hall effect for photons \cite{Leykam2018,Mittal2019}, with a Haldane-like tight-binding Hamiltonian
\begin{eqnarray}
H_{\text{L}} &=& \sum_{i,j} \W_{0} a^{\dag}_{i} a_{i}  \\
  &-& J \left( \sum_{\left<i,j \right>} a^{\dag}_{j} a_{i} e^{- i \phi_{i,j}} + \sum_{\left<\left< i,j \right>\right>} a^{\dag}_{j} a_{i} + \mathrm{h.c.} \right) \nonumber
\end{eqnarray}
Here $a^{\dag}_{i} \left(a_{i}\right)$ is the creation (annihilation) operator at a lattice site $i = \left(x,y \right)$. The summations $\left<i,j \right>$ and $\left<\left< i,j \right>\right>$ are over the nearest and the next-nearest neighbor lattice sites, respectively. $J$ is the coupling strength between the lattice sites, and is the same for both nearest and next nearest neighbors. The hopping phase $\phi_{i,j} = \pm \pi/4$ for nearest-neighbor couplings, and $\phi_{i,j} = 0$ for next-nearest neighbor couplings. The energy-momentum band structure of the lattice exhibits two bulk bands separated by a bandgap. For a finite lattice, the band gap hosts topological edge states that are confined to the boundary of the lattice. More importantly, the edge states are robust against disorders, such as, a mismatch in the ring resonance frequencies. Furthermore, they exhibit a linear dispersion \cite{Leykam2018, Mittal2019, Mittal2014}. The band structure of the lattice and the presence of edge states can be probed by measuring the transmission and the delay spectra of the lattice from input to the output port. The simulated transmission and delay spectra for a $8 \times 8$ lattice are shown in Figs.\ref{fig:1}(b,c). The linear dispersion of the edge states manifests in the Wigner delay spectrum as a flat profile (Fig.\ref{fig:1}(c)) in the range $\dW = [-1, 1] J$, where $\dW = \omega - \omega_{0}$ is the detuning of the excitation laser frequency $\left(\omega \right)$ from the ring resonance frequency $\left(\omega_{0} \right)$, for a given longitudinal mode. In contrast, the Wigner delay in the  bulk band varies significantly because in a finite lattice, the bulk bands do not have a well defined momentum. Our system also supports a pseudospin degree of freedom because of the two circulation directions (clockwise and counter-clockwise) in the ring resonators. The two pseudospins (up and down) are time-reversed partners, and therefore, experience opposite hopping phases and exhibit counter propagating edge states \cite{Mittal2019}.

Our devices are fabricated using the CMOS compatible silicon-on-insulator (SOI) platform at a commercial foundry (IMEC Belgium). The ring waveguides are $\simeq$ 510 nm wide, $\simeq$ 220 nm high, and at telecom wavelengths ($\approx$ 1550 nm), support a single TE polarized mode. The ring length is $\simeq 70 \um$ with a free-spectral range of $\approx$ 1 THz. The coupling gap between the rings is 0.180 nm, and it results in a coupling strength $J \simeq (2\pi)15.6$ GHz. The lattice is coupled to input and output waveguides as shown in Fig.\ref{fig:1}. At the ends of the input/output waveguides, we use grating couplers to inject light from standard single-mode fiber into the waveguides. Figure \ref{fig:2}(a) shows the measured transmission spectrum of the device for spin-up excitation, with the edge band highlighted in color. The edge states for this excitation take the shorter route from input to the output coupler, as shown in Fig.\ref{fig:1}(a).

To generate indistinguishable photon pairs in this lattice, we use the $\chi^{(\text{3})}$ nonlinearity of silicon and implement a dual-pump (DP) SFWM process (Fig.\ref{fig:1}(a)). We pump the lattice using two classical, continuous-wave pump beams, at frequencies $\W_{p_{1}} \text{and} ~\W_{p_{2}}$. The DP-SFWM then leads to the generation of correlated photon pairs, called signal and idler, at frequencies $\W_{s} \text{and} ~\W_{i}$, respectively, such that the energy conservation relation $\W_{p_{1}} + \W_{p_{2}} = \W_{s} + \W_{i}$ is satisfied. This nonlinear process is described by the Hamiltonian
\begin{eqnarray}
H_{\text{NL}} &=&  ~\eta \sum_{m}  \Bigl( a^{\dag}_{m,s} ~ a^{\dag}_{m,i} ~ a_{m,p_{1}} ~a_{m,p_{2}} \\
 &+& a^{\dag}_{m,p_{1}} ~a^{\dag}_{m,p_{2}} ~a_{m,s} ~a_{m,i} \Bigr).\nonumber
\end{eqnarray}
Here $\eta$ is the efficiency of the SFWM process, and $a^{\dag}_{m,\mu}$, is the photon creation operator for signal, idler or pump photons at a lattice site $m$.

The two-photon state generated at the output of our device is described, in general, as
\begin{equation}
\left|\Psi \right> = \int d\W_{s} d\W_{i} \phi\left(\W_{s},\W_{i}\right) \delta\left(\W_{s} + \W_{i} - \W_{p_{1}} - \W_{p_{2}}\right) a^{\dag}_{s} a^{\dag}_{i} \left|vac \right>,
\end{equation}
where $a^{\dag}_{s,i}$ are the photon creation operators for the signal or idler photons, and $\phi\left(\W_{s},\W_{i}\right)$ is the spectral amplitude of the two-photon wavefunction. Given the fact that our ring resonator waveguides support a single TE polarized mode and the generated photons are collected from a single spatial mode (the same output port), the generated photons are indistinguishable in all degrees of freedom when the two-photon spectral amplitude is symmetric with respect to exchange of photons, that is, $\phi\left(\W_{s},\W_{i}\right) = \phi\left(\W_{i},\W_{s}\right)$ \cite{Branning2000, Harder2013}. Furthermore, the photon pairs are energy-time entangled when $\phi\left(\W_{s},\W_{i}\right) \neq \phi_{s}\left(\W_{s}\right) \phi_{i}\left(\W_{i}\right)$, that is, when the two-photon spectral wavefunction can not be expressed as a product of individual wavefunctions of signal and idler photons \cite{Harder2013}.

In our experiment, we position the two pump frequencies in two different longitudinal modes of the lattice separated by two free-spectral ranges (FSR, $\Omega$, see Fig.\ref{fig:1}(b)). The indistinguishable photon pairs are then generated in the longitudinal mode located midway between the two pump modes, that is, $\W_{0p_{1}} + \W_{0p_{2}} = \W_{0s} + \W_{0i}$ and $\W_{0s} = \W_{0i}$. Here $\W_{0\mu}$, with $\mu =  p_1, p_2, s, i$, is the resonance frequency of the respective longitudinal mode. We note that each of the two pump beams also generate distinguishable photon pairs via non-degenerate (single pump) SFWM (SP-SFWM). However, because of the energy conservation, these photon pairs are generated in longitudinal modes located symmetrically around the respective pump beams  (Fig.\ref{fig:1}(b)). Therefore, we use spectral filtering and time-resolved coincidence measurements between detected photon pairs to exclude the noise photons generated by single pump SFWM  (Fig.\ref{fig:1}(c)).

To understand the nature of spectral correlations between the two pump fields and the generated photons, we first measure the number of indistinguishable photon pairs generated via DP-SFWM as a function of the two pump frequency detunings $\left(\dW_{p_{1,2}} = \W_{p_{1,2}} - \W_{0p_{1,2}}\right)$, relative to their respective longitudinal mode center frequencies. As mentioned earlier, we use time-resolved correlation measurements $\left(g^{\left(2\right)}\left( \tau\right) \right)$ to post-select the photon pairs generated by DP-SFWM. Figure \ref{fig:2}(b) shows the typical temporal correlation function with pump powers $P_{1} = 1$ mW and $P_{2} = 3$ mW at the input of the lattice. We measure a maximum $g^{\left(2\right)}\left(0\right) \simeq 117$ which shows that the two photons are indeed correlated. We integrate over the correlation peak to get the total number of coincidence counts in a given acquisition time (here 10 seconds). Figure \ref{fig:2}(c) shows the measured number of coincidence counts as a function of the frequency detunings $\dW_{p_{1}}, \dW_{p_{2}}$. We observe that the photon generation rate is maximum when both the pump frequencies are in the edge band, that is, when $\dW_{p_{1}}, \dW_{p_{2}} = [-1, 1] J$ (Fig.\ref{fig:2}(a)). Furthermore, compared to the bulk band regions, the generation rate is relatively uniform throughout the edge band.  We note that for a given choice of the two pump frequencies, energy and momentum conservation lead to spectral correlations between generated photons. However, our measurement of the number of generated photon pairs as a function of the pump frequencies does not resolve these spectral correlations.

To reveal the spectral correlations between generated signal and idler photons, we fix the input pump frequencies to be in the middle of the edge band, at $\dW_{p_{1}} \simeq 0 \simeq ~\dW_{p_{2}}$, and measure the joint-spectral intensity (JSI), $ \left|\phi\left(\dW_{s},\dW_{i}\right)\right|^{2}$ (Fig.\ref{fig:2}(d)). This is the joint probability of detecting a signal photon at frequency $\dW_{s}$ and an idler photon at frequency $\dW_{i}$. Here $\dW_{s,i}$ are the frequency detunings of the signal and idler photons relative to their respective longitudinal mode resonances. The measured correlations show that, with the two pump fields in the edge band, the spectrum of generated signal and idler photons is also limited to the edge band. This is because of the linear dispersion of the edge states that leads to efficient phase matching (momentum conservation) when all the four fields are in the edge band, and confinement of the edge states to the lattice boundary that leads to a good spatial overlap between the fields. Furthermore, both the signal and idler spectra are centered around $\dW_{s} \simeq 0 \simeq \dW_{i}$ which shows that they are degenerate in frequency, that is, $\left| \phi\left(\W_{s},\W_{i}\right) \right|^{2} = \left| \phi\left(\W_{i},\W_{s}\right) \right|^{2}$. The JSI also shows that the signal and idler photons generated by our source are entangled, that is, $\phi\left(\W_{s},\W_{i}\right) \neq \phi_{s}\left(\W_{s}\right) \phi_{i}\left(\W_{i}\right)$. Note that we use continuous-wave pumps in our experiments, and the apparent width of spectral correlations along the diagonal is because of the finite spectral resolution $\left(\approx 10 ~\text{GHz} \simeq 0.64 J \right)$ of our measurements.

The energy conservation and the linear dispersion of the edge states allows us to tune the spectral bandwidth of generated photons by tuning the input pump frequencies within the edge band region. This is because of the efficient momentum conservation in the edge band that limits the spectra of generated photons also to the edge band region. To show such tunability of the spectra of generated photons, we measure the signal-idler spectral correlations for different pump frequencies in the edge band (Fig.\ref{fig:2}(e,f)). When both the pump frequencies are near the side of the edge band $\left(\simeq 0.8 J \right)$, in Fig.\ref{fig:2}(e), we observe that the spectra of generated photons are significantly narrower $\left(\text{by} \approx 4\times \right)$ than that when both the pumps are in the center of the edge band (Fig.\ref{fig:2}(d)). Also, the spectra are centered around $0.8J$ which shows that the two photons are degenerate in frequency, as expected. Similarly, when the two pump frequencies are at different locations in the edge band $\left(\delta\W_{p_{1}} \simeq 0.8 J, \dW_{p_{2}} \simeq 0 \right)$, we observe that the spectra of generated photons are centered around $0.4 J$, with a bandwidth larger than that with both the pumps in the side of the edge band (Fig.\ref{fig:2}(f)).

\begin{figure*}
\includegraphics[width=0.9\textwidth]{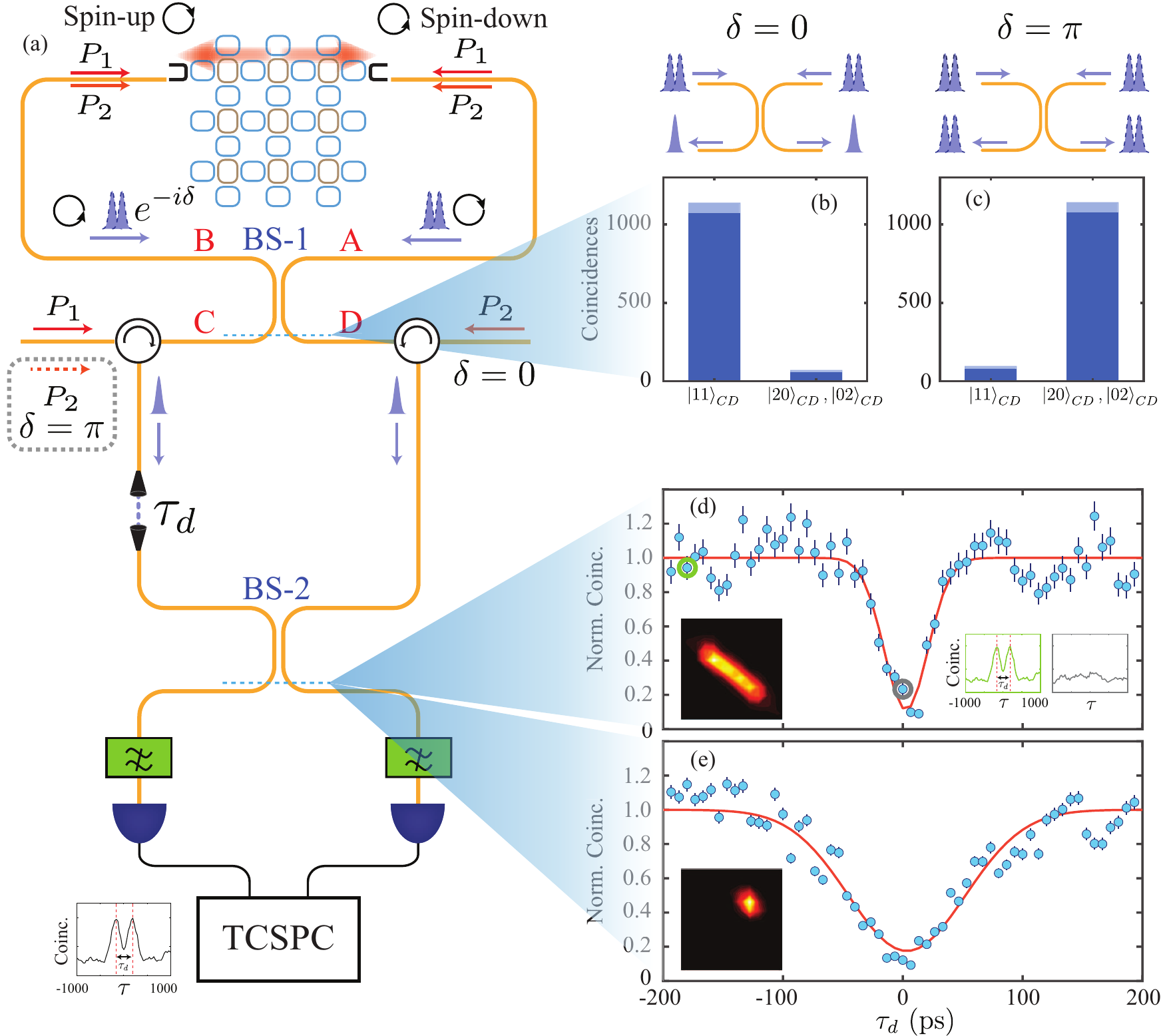}
\centering
\caption{
(a) Schematic of the Sagnac interferometer setup used to deterministically split the two photons via time-reversed HOM interference of a path-entangled two-photon state (at beamsplitter BS-1), and subsequently, realize HOM interference at beamsplitter BS-2 with a variable delay $\tau_{d}$ between the two photons. We used time-resolved coincidence measurements at the two detectors. (b,c) Measured two-fold coincidences at the output ports C, D of the beamsplitter BS-1, for two different configurations of the input pump beams. The photons anti-bunch ($\delta = 0$) when the two pumps are in separate input ports of BS-1, and the photons bunch ($\delta = \pi$) when the pumps are in the same input port. (d,e) Measured HOM interference dip for $\delta = \pi$, and pump frequencies $\left(\dW_{p_1},\dW_{p_2} \right)$ = $\left(0, 0 \right) J$, and $\left(0.8, 0.8 \right) J$, respectively. Insets show measured JSI and time-resolved coincidences between the two photons.
}
\label{fig:3}
\end{figure*}

Though our joint-spectral intensity measurements show that the signal and idler photons are degenerate in frequency, $\left| \phi\left(\W_{s},\W_{i}\right)\right|^2 = \left|\phi\left(\W_{i},\W_{s}\right) \right|^{2}$, these measurements do not confirm their indistinguishability which requires phase-coherence such that $\phi\left(\W_{s},\W_{i}\right) = \phi\left(\W_{i},\W_{s}\right)$ \cite{Branning2000, Harder2013}. The way to unambiguously confirm the indistinguishability of generated signal and idler photons is to perform Hong-Ou-Mandel (HOM) interference \cite{Hong1987} between the two photons. In HOM interference, when two indistinguishable photons arrive simultaneously at the two input ports of a beamsplitter, they bunch together at the output of the beamsplitter. We emphasize that HOM interference between correlated signal and idler photons only requires the two-photon spectral wavefunction to be symmetric, $\phi\left(\W_{s},\W_{i}\right) = \phi\left(\W_{i},\W_{s}\right)$, but not necessarily separable, $\phi\left(\W_{s},\W_{i}\right) = \phi_{s}\left(\W_{s}\right) \phi_{i}\left(\W_{s}\right)$ \cite{Branning2000, Harder2013}.

In HOM interference the two photons arrive separately, one photon in each of the two input ports of the beamsplitter. However, in our topological source, both the photons are in a single spatial mode, they have the same polarization, and they are degenerate in frequency. Therefore, we can not deterministically split the two photons into two spatial modes using a normal beamsplitter which creates at its output a superposition of states where one photon is in each port, or two photons are in the same port \cite{Chen2007, He2015}. Nevertheless, when the input to the beamsplitter is a path-entangled two-photon state of the form $\left|2 0 \right>_{A,B} + \left|0 2 \right>_{A,B}$, that is, when both the photons arrive either at the input port A or at port B of the beamsplitter, then the two-photon state at the output ports C, D of the beamsplitter is deterministic with one photon in each port, that is, $\left|1 1 \right>_{C,D}$ \cite{Chen2007, He2015}. Here the state $\left|n m \right>_{A(C),B(D)}$ refers to $n$ photons in the input(output) port A(C) of the beamsplitter and $m$ photons in the input(Output) port B(D). This scenario, in fact, corresponds to time-reversed HOM interference of two photons \cite{Chen2007}.

To deterministically split the two photons, so that we can later perform HOM interference between them, we use our topological source in a Sagnac interferometer (formed by beamsplitter BS-1, Fig.\ref{fig:3}(a)) \cite{Chen2007, He2015}. In this configuration, both the pseudospins (up and down) associated with our source are simultaneously pumped. Because they are time-reversed partners, the pump beams corresponding to the two pseudospins propagate through the same edge state, but in opposite directions, and generate an entangled two-photon state $\left|2 0 \right>_{A,B} + e^{-i\delta} \left|0 2 \right>_{A,B}$ at ports A, B of the beamsplitter BS-1 (Fig.\ref{fig:3}(a)). We note that the strength of SFWM interaction in our experiment is very weak such that the probability of generating two photon pairs, one in each arm of the Sagnac interferometer, is small. The relative phase $\delta$ of two-photon entangled state can be set to $0$ or $\pi$ by appropriately choosing the input ports for the two pump beams at the Sagnac beamsplitter (BS-1 in Fig.\ref{fig:3}(a)) \cite{He2015}. When both the pumps are in the same port of the BS-1 (Port C or Port D), the phase $\delta = \pi$, and the two photons bunch at the output of  BS-1, that is, they appear together at either port C or port D of BS-1 (Fig.\ref{fig:3}(c)). In contrast, when the two pumps are in different ports of the beamsplitter BS-1 (one in Port C, and the other in Port D), the phase $\delta = 0$ and it leads to anti-bunching of photons such that the photons are deterministically separated at the output of the BS-1 (Fig.\ref{fig:3}(b)). We use two circulators to collect the photons at ports C and D. For $\delta = 0$, we measure the total probability of bunching (in either port C or port D), $g^{\left( 2\right)} \left(0\right) = 0.05(1)$, which shows that the two photons are predominantly in the state $\left|1 1 \right>_{C,D}$. For $\delta = \pi$, we measure $g^{\left( 2\right)} \left(0\right) = 0.93(1)$, which shows that the two photons are still in the same spatial mode (port C or port D). We emphasize that the use of a Sagnac interferometer, with the two pump beams injected at different input ports, alleviates the need for any active stabilization of our source.

To demonstrate HOM interference we set $\delta = 0$ such that the two photons are deterministically separated in the ports C and D of the beamsplitter BS-1. We pump our source in the middle of the edge band, that is, $\delta\W_{p_{1}} \simeq 0 \simeq \delta\W_{p_{2}}$. We introduce a relative delay $\left(\tau_{d} \right)$ between the two photons, interfere them on another beamsplitter (BS-2), and measure the coincidence counts at the output of BS-2 as we vary the delay $\tau_{d}$ (Fig.\ref{fig:3}(a)). We see a HOM dip in the coincidence counts, with a visibility of $88 (10)\%$, which confirms that the two photons are indeed indistinguishable (Fig.\ref{fig:3}(d)).

We note that the temporal width of the HOM interference dip is inversely related to the spectral width of the joint-spectral intensity (along the line $\delta\W_{s} = -\delta\W_{i}$) that characterizes the two-photon state. As we demonstrated in Fig.\ref{fig:2}(d-f), we can control the JSI of generated photons in our source by simply tuning the input pump frequencies (Fig.\ref{fig:2}). To demonstrate similar control in the HOM interference, we set the two pump frequencies to be at one of the extremes of the edge band $\delta\W_{p_{1}} \simeq 0.8 J \simeq \delta\W_{p_{2}}$ such that the spectral width of the JSI is limited to $\approx 0.8J$ (Fig.\ref{fig:2}d). We now observe, in Fig.\ref{fig:3}(e), that the temporal width of the HOM interference dip is indeed much larger (by a factor of $\approx$ 2.7(4)) compared to the case with both the pumps in the center of the edge band. The discrepancy between this factor and the decrease in the spectral width (by a factor of 4.0(6)) can be accounted for by the limited spectral resolution of our JSI measurement.

\begin{figure}
\includegraphics[width=0.46\textwidth]{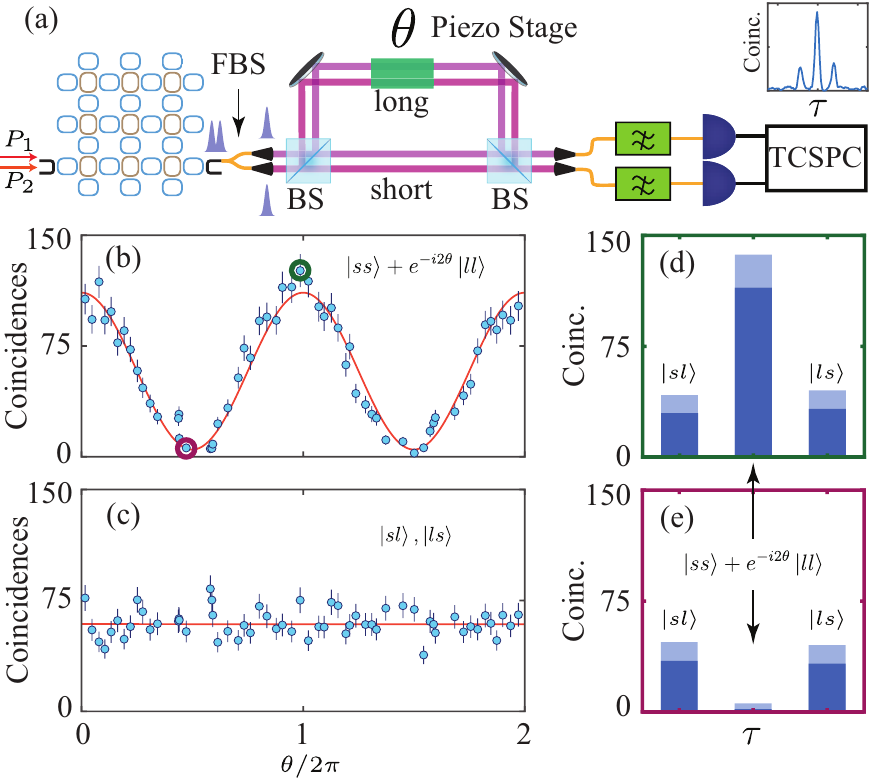}
\caption{
(a) Schematic of the Franson interferometer setup to show the energy-time entanglement of generated photon pairs. The delay between the short and the long paths is $\approx 800$ps, much longer than the temporal correlation between the two photons. (b) Measured two-fold coincidences in the central peak, that is, when both the photons travel through the same path (short or long), as a function of the interferometer phase $\theta$. (c) Measured two-fold coincidences in the side peaks, that is, when the photons travel through different paths and are therefore, distinguishable. (d-e) Measured histograms at two points indicated in (b), showing the maximum and the minimum coincidence counts in the central peak. The counts in the side peaks are almost constant. The error bars are calculated assuming Poisson statistics. For this measurement the input pump frequencies were set to $\delta\W_{p_{1}} \simeq 0.8 J \simeq \delta\W_{p_{2}}$.
}
\label{fig:4}
\end{figure}

Finally, we show that the generated two-photon state is energy-time entangled. The use of a continuous-wave pump for generating photon pairs via SFWM (or SPDC) naturally leads to the emergence of energy-time entanglement such that $\Delta\W \Delta T < 1$, where $\Delta\W = \Delta\left(\W_{s} + \W_{i}\right)$ is the uncertainty in the total energy of the signal and idler photons, set by the pump bandwidth. $\Delta T$ is the uncertainty in the time duration between arrival of two photons \cite{Franson1989}. We use a beam splitter to probabilistically split the two photons at the output of our source, and inject them into two Franson interferometers with a path-length delay ($\approx 800$ ps) much larger than the temporal correlation of the generated photons ($\Delta T \approx 200$ ps, as shown in Fig.\ref{fig:2}(b)). Time-resolved coincidence measurements at the outputs of the interferometers show three peaks (Fig.\ref{fig:4}). The two  side peaks correspond to the two cases when one of the photons took a shorter path in the interferometer, and the other took a longer path. The center peak corresponds to the two cases when both photons took either the shorter path ($\left|ss\right>$) or the longer path ($\left|ll\right>$). We measure the number of coincidence counts in the three peaks as we vary the phase $\theta$ of the interferometers. The latter two cases (both short or both long) are indistinguishable, and therefore, we observe interference fringes in the coincidence counts as a function of the two-photon phase $2\theta$ acquired in the interferometer. In contrast, the other two cases, where one photon travels through the shorter path and the other through the longer path ($\left|sl\right>,\left|ls\right>$), are distinguishable and accordingly yield no interference. Our observation of two-photon interference fringes, for a path-length delay that is much longer than the temporal correlation width $\Delta T$ of the photons, demonstrates that the coherence time of the generated two-photon state $\left(1/\Delta\W \right)$ is much longer than $\Delta T$ and the two-photon state is indeed energy-time entangled.

Because the edge states are topologically protected, we expect that the spectral correlations between generated photon pairs will also be robust against fabrication disorders when the two pump frequencies, and therefore, the signal and idler frequencies are in the edge band. Indeed, in ref.\cite{Mittal2018}, we demonstrated the topological robustness of spectral correlations using a single-pump SFWM process \cite{Mittal2018}. To show that this topological robustness holds for dual-pump SFWM process as well, we provide numerical simulation results in the Supplementary Information (section S5). We fix the input pump frequencies to be in the center of the edge band and calculate the spectra of generated photons for random realizations of disorder. We compare these results against those for a 1D array of ring resonators which is topologically trivial, and therefore, not expected to be robust against disorder. As expected, we observe that our topological sources of indistinguishable photon pairs achieves much higher spectral similarity across devices when compared to topologically trivial sources.

In summary, we have demonstrated a topological source of indistinguishable photon pairs with tunable spectral-temporal correlations. Our demonstration could lead to on-chip generation of novel quantum states of light where topological phenomena are used for robust manipulations of the photonic mode structure and quantum correlations between photons. In particular, in the low-loss regime, our topological device can be used to generate single-mode squeezed light \cite{Peano2016, Vernon2019, Zhang2020, Zhao2020} with a tunable spectrum, and at the same time, overcome the gain-bandwidth product that is inherent to single ring resonators. On a more fundamental level, nonlinear parametric processes such as four-wave mixing are inherently non-Hermitian in nature, that is, they do not conserve particle number. Therefore, our system paves the way for investigations of the rich interplay between topology, non-Hermitian physics, and quantum photonics processes to realize novel topological phases that are unique to photons. \\

\noindent
\textbf{Acknowledgements:} This research was supported by the Air Force Office of Scientific Research AFOSR-MURI grant FA9550-16-1-0323, Office of Naval Research ONR-MURI grant N00014-20-1-2325, Army Research Laboratory grant W911NF1920181, and NSF grant PHY1820938. We thank Qudsia Quraishi for the nanowire detectors.\\

\bibliographystyle{NatureMag}
\bibliography{IFWM_Biblio}

\end{document}



\title{Tunable quantum interference using a topological source of indistinguishable photon pairs}

\author{Sunil Mittal$^{1,\,2,\,*,\,\dag}$}
\author{Venkata Vikram Orre$^{1,\,2,\,*}$}
\author{Elizabeth A. Goldschmidt$^{3}$}
\author{Mohammad Hafezi$^{1,\,2,\,4}$}

\affiliation{$^1$Joint Quantum Institute, NIST/University of Maryland, College Park, Maryland 20742, USA}
\affiliation{$^2$Department of Electrical and Computer Engineering and IREAP, University of Maryland, College Park, Maryland 20742, USA}
\affiliation{$^3$Department of Physics, University of Illinois at Urbana-Champaign, Urbana, Illinois 61801, USA}
\affiliation{$^4$Department of Physics, University of Maryland, College Park, MD 20742, USA}
\affiliation{$^*$S.M. and V.V.O. contributed equally}
\affiliation{$^\dag$Email: mittals@umd.edu}

\maketitle

\section{S1: Coincidences and CAR as a function of pump power}

\begin{figure}[h]
\centering
\includegraphics[width=0.98\textwidth]{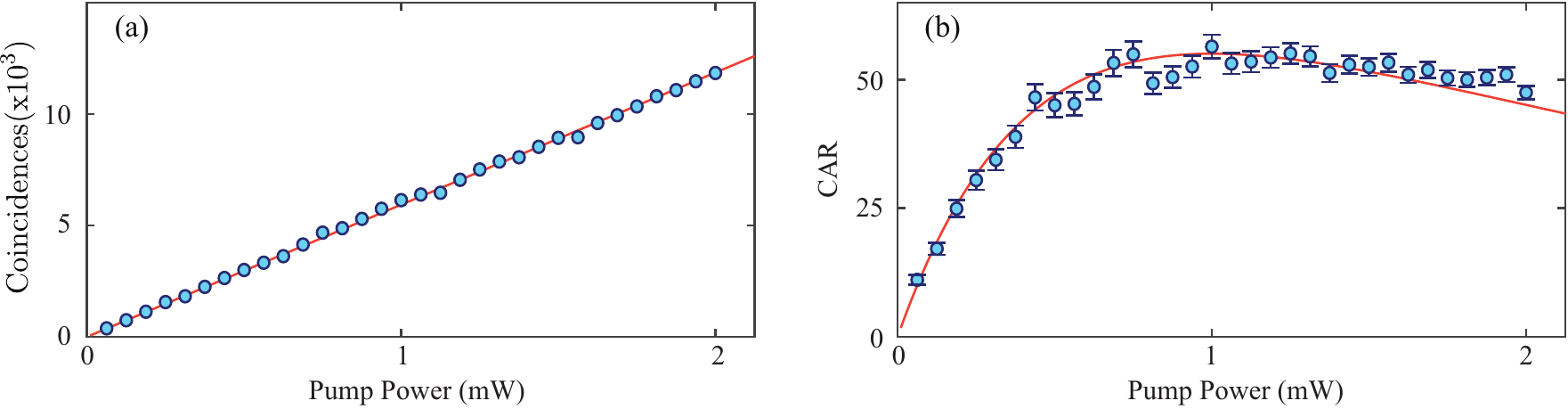}
\caption{(a) Coincidences, and (b) CAR as function of one of the pumps powers, while the other pump power is fixed at 3 mW.}
\label{fig:S1}
\end{figure}

The number of coincidence counts generated in the dual-pump spontaneous four-wave mixing (DP-FWM)  process depend on the product of the input pump powers as $P_{p_1}P_{p_2}$, where $P_{p_{i}}$ is the power of the pump with index $i = 1, 2$. In contrast, the number of coincidence counts generated in a single-pump spontaneous four-wave mixing (SP-FWM) process vary as $P_{p_i}^{2}$. To verify that the coincidence counts detected in our setup are generated by the DP-SFWM process, we fix one of the input pump powers, at 3 mW, and measure the coincidence counts as we vary the other pump power. From Fig.\ref{fig:S1}(a), we see that the number of coincidence counts indeed increase linearly with one of the pump powers.

We also measure the coincidences-to-accidentals ratio (CAR) which is an indication of the signal to noise ratio of a source. To do that, we integrate the measured $g^{\left(2\right)}(\tau)$ (see Fig.2(b) of the main text) around the peak to calculate the actual coincidence counts, and divide it by the mean coincidence counts at $\tau \gg 0$ which are the accidental coincidences. Our source achieves a maximum CAR of $\approx$ 53. We also measured CAR as we vary one of the pump powers (say $P_{p_2}$) while the other pump power (say $P_{p_{1}}$) is fixed (at 3 mW), and the result is shown in Fig.\ref{fig:S1}(b). As expected, we find that the CAR follows the relation \cite{Silverstone2013}
\begin{equation}
CAR = \frac{\eta P_{p_1}P_{p_2}}{(P_{p_1}^2 + P_{p_2}^2+ P_{p_1}P_{p_2})^2},
\end{equation}
where $\eta$ is a fitting parameter. For low $P_{p_2}$, the CAR is limited by the noise photons generated by pump $p_{1}$ via single-pump SFWM process.

\section{S2: JSI in the bulk band}

\begin{figure}
\centering
\includegraphics[width=0.78\textwidth]{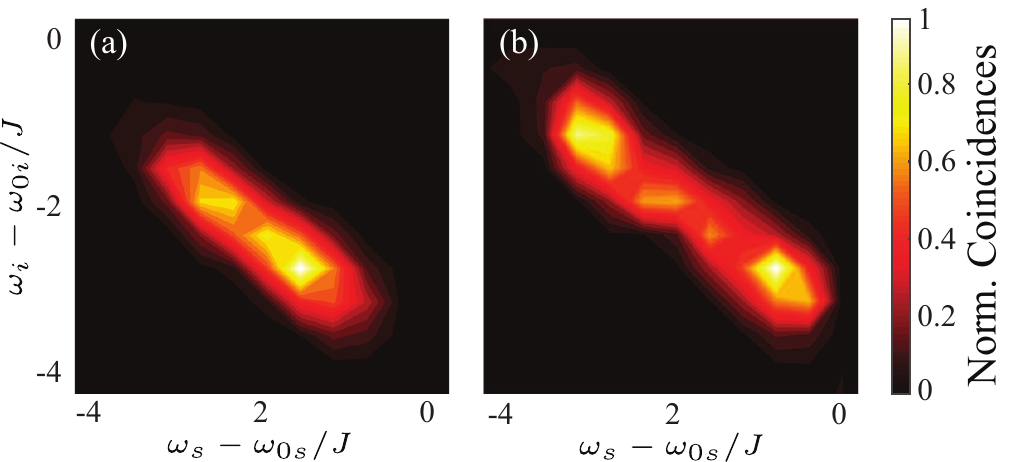}
\caption{ Measured JSI when the input pumps frequencies are in the bulk band, (a) $\left(\delta\omega_{p_1}, \delta\omega_{p_2} \right) = \left(-2.42,-2.42 \right) J$, (b) $\left(\delta\omega_{p_1}, \delta\omega_{p_2} \right) = \left(-1.82,-2.42\right) J $. The JSI shows presence of distinct modes, that change with pump frequencies.
}
\label{fig:S2}
\end{figure}

Figure ~\ref{fig:S2} shows the measured the joint-spectral intensity (JSI) of the signal and idler photons, when the input pump frequencies are in the bulk band of the device, that is, when $\left(\delta\omega_{p_1}, \delta\omega_{p_2} \right) = \left(-2.42,-2.42 \right) J, \text{and} \left(-1.82,-2.42\right) J$. While the JSI is still along the anti-diagonal $\omega_{s} = -\omega_{i}$ because of energy conservation, we observe that the JSI shows presence of distinct modes. More importantly, this mode pattern changes randomly as we tune the input pump frequencies. In contrast, when both the pump frequencies are in the edge band (Fig. 2 of the main text), the JSI is relatively uniform, and only the bandwidth of the JSI changes as we change the pump frequencies. This behavior of bulk band is because, unlike the edge band, it does not have a well-defined momentum and its dispersion is very sensitive to fabrication disorders.

\section{S3: Franson interferometer for demonstrating energy-time entanglement}

\begin{figure}
\centering
\includegraphics[width=0.98\textwidth]{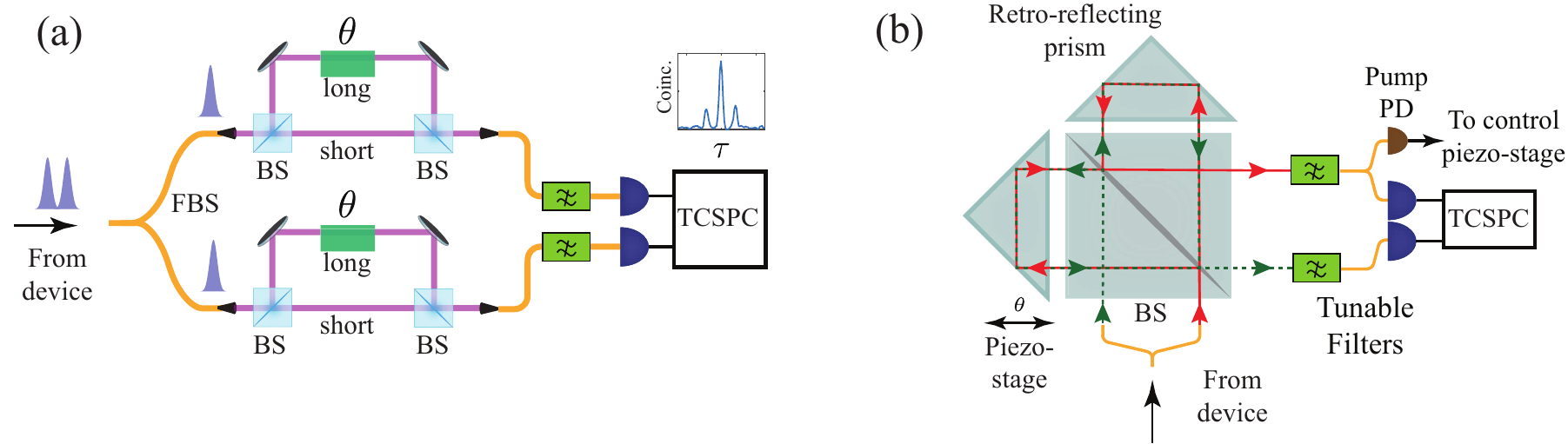}
\caption{(a) Simplified schematic of the Franson setup used to demonstrate energy-time entanglement between between generated photon pairs. (b) Schematic of the actual Franson interferometer setup implemented using Michelson configuration. FBS: fiber beamsplitter, BS: beamsplitter, PD: photo-detector, TCSPC: time-correlated single photon counter}
\label{fig:S3}
\end{figure}

In figure 4 of the main text, we showed a simplified schematic of the Franson interferometers that were used demonstrate energy-time entanglement between generated signal and idler photon pairs. In our experiment, for the ease of alignment, we actually used a single Michelson interferometer to implement both the unbalanced interferometers as shown in Figure \ref{fig:S3}. In this configuration, the relative phase $\theta$ of both the interferometers is the same. We used a piezo-controlled stage to change the phase $\theta$. Furthermore, we used WDM filters to filter out the pump wavelength, measured the pump power using a photodetector (PD), and used this measurement to monitor the phase $\theta$ of the interferometer.

\section{S4: Robustness of spectral correlations}

\begin{figure}
\centering
\includegraphics[width=0.5\textwidth]{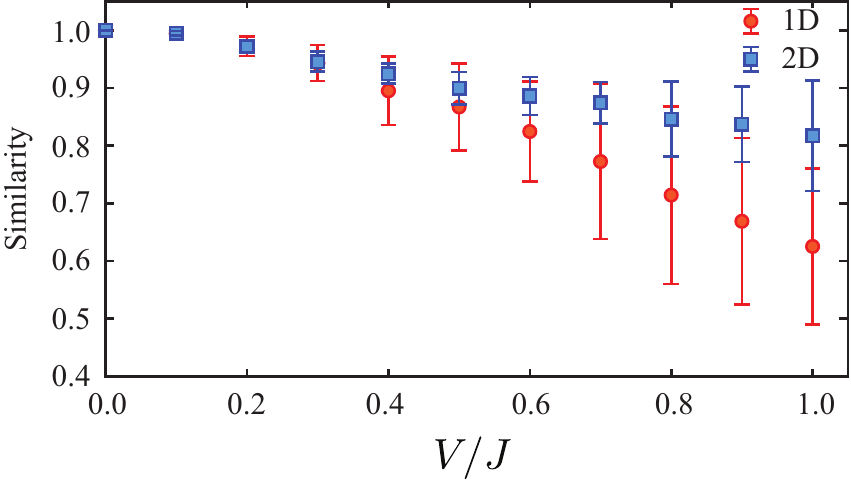}
\caption{Numerical simulation results comparing the similarity in the spectra of generated photons as a function of disorder for topologically trivial 1D array and 2D topological devices. As expected the spectra of generated photons in topologically trivial 1D devices are much more susceptible to disorder compared to the 2D topological devices. The simulation data was acquired for fixed input pump frequencies $\left(\omega_{p_{1}},\omega_{p_{2}} \right) = (0,0)J$.}
\label{fig:S4}
\end{figure}

To show the robustness of spectral correlations of our source in the presence of disorder, we perform numerical simulations to calculate the spectra of generated photons in the presence of random disorder. We introduce the disorder using random variations in the ring resonator frequencies, where the disorder strength is given by $V$. We fix the input pump frequencies $\left(\omega_{p_{1}},\omega_{p_{2}} \right) = (0,0)J$, and average over 100 random realizations for each disorder. We then calculate the mean similarity of the disordered spectra with the spectra in the absence of any disorder. The similarity between two devices $i$ and $j$ is calculated by taking the mean of the inner product between two spectra and is given by,
\begin{equation}
S_{i,j} = \frac{\sqrt{\int\Gamma_{i} \Gamma_{j}}^2}{\int\Gamma_{i}\int\Gamma_{j}}
\end{equation}
where $\Gamma_{i,j}$ is the spectra of generated photons for a device $i$,$j$. We compare the results against those obtained for a topologically trivial 1D array of 10 ring resonators. As shown in Fig.\ref{fig:S4}, our topological source of indistinguishable photon pairs achieves much higher spectral similarity across devices when compared to a topologically trivial source. We also observe that the variability in the spectra of 1D devices is much higher compared to 2D devices for larger disorders. These results show the advantage of using topologically robust edge states for generating indistinguishable photon pairs \cite{Mittal2018}.

\bibliographystyle{NatureMag}
\bibliography{IFWM_Biblio}